\documentclass[a4paper,11pt]{article}
\pdfoutput=1 

\usepackage{jheppub} 
\usepackage{natbib}
\usepackage[T1]{fontenc} % if needed
\usepackage{amsmath}
\usepackage{slashed}
\DeclareMathOperator{\sign}{\text{ sign}}

\title{A test of bosonization at the level of four-point functions in Chern-Simons vector models}

\author{Akshay Bedhotiya}
\author[a]{and Shiroman Prakash}

\affiliation[a]{Dayalbagh Educational Institute, Agra, India}

\emailAdd{aksbed@gmail.com}
\emailAdd{shiroman@gmail.com}

\abstract{We study four-point functions in Chern-Simons vector models in the large $N$ limit. We compute the four-point function of the scalar primary to all orders in the `t Hooft coupling
$\lambda=N/k$ in $U(N)_k$ Chern-Simons
theory coupled to a fundamental fermion, in both the critical and non-critical theory, for a particular case of the external momenta. These theories cover the entire 3-parameter "quasi-boson" and 2-parameter "quasi-fermion" families of 3-dimensional quantum field theories with a slightly-broken higher spin symmetry. Our results are consistent with the celebrated bosonization duality, as we explicitly verify by calculating four-point functions in the free critical and non-critical bosonic theories. }

\begin{document} 
\maketitle
\flushbottom

\section{Introduction}
$U(N)_k$ Chern-Simons theories coupled to fundamental matter provide an interesting class of interacting three-dimensional conformal field theories that are exactly solvable in the 't Hooft limit: $N \rightarrow \infty$, $k \rightarrow \infty$ with the 't Hooft coupling $\lambda \equiv \frac{N}{k}$ held fixed. These theories, which have been intensively studied in the past few years \cite{Giombi:2011kc,  Aharony:2011jz, Aharony:2012nh,  GurAri:2012is, Jain:2013py-2, Jain:2014nza, Jain:2013py-1,Aharony:2012ns,Jain:2013gza,Jain:2012ss,Minwalla:2011ma,Yokoyama:2012fa,Minwalla:2015new}, are particularly interesting because they provide examples of non-supersymmetric dualities. For instance, they are widely believed to be dual to Vasiliev higher-spin gauge theories (see \cite{ Giombi:2012ms} for a review.) They also exhibit a spectacular bosonization duality relating Chern-Simons theory coupled to fundamental fermions to critical Chern-Simons theory coupled to fundamental bosons, which can be thought of as a non-supersymmetric generalisation of the ABJ and Giveon-Kutasov dualities. \cite{Giombi:2011kc,Aharony:2011jz, Aharony:2012nh,  GurAri:2012is, Jain:2013py-2, Jain:2014nza, Jain:2013py-1,Aharony:2012ns,Jain:2012ss, Chang:2012kt, Giveon:2008zn,Benini:1}

The bosonization duality has been tested via three point functions and also in thermal free energy computations, leaving little doubt to its correctness. However, it is still of independent interest to directly test the duality at the level of four-point functions; which are not determined by purely kinematic considerations.  

In this paper, we calculate four-point correlation functions of  the primary scalar operator $J^{(0)}$ in the critical and non-critical $U(N)_k$ Chern-Simons theory coupled to fundamental fermions. For a particular choice of external momenta, we are able to obtain a closed form (but highly non-trivial) expression for the four point function of the scalar primary as a function of $\lambda$ -- which we then compare to the free and critical bosonic theories to obtain another independent check of the bosonization duality.

\label{sec:intro}

\subsection{Review of the Bosonization Duality}

The bosonization duality \cite{Aharony:2012nh}, which can be thought of as a non-supersymmetric generalization of the Giveon-Kutasov duality \cite{Giveon:2008zn}, states that a $U(N_f)_{k_f}$ Chern-Simons theory coupled to fermions in the fundamental representation is dual to a $U(N_b)_{k_b}$ Chern-Simons theory coupled to critical bosons in the fundamental representation. The critical theory is obtained by deforming the usual (non-critical) theory by a double trace operator $\lambda_4 \phi^2 \phi^2$ and taking the coupling to infinity. (The coupling $\lambda_4$ should not be confused with $\lambda_b=N_b/k_b$.)

The conjectured duality claims that the two theories are equivalent, with the following relation between parameters:
\begin{align}
k_f  =  -k_b \\
N_f  =  |k_b| - N_b
\end{align}
Though we present the duality in terms of $k$ and $N$, the duality has only been tested in the large $N$, 't Hooft limit; at finite $N$ there will be some shifts of $\pm 1/2$ in the Chern-Simons level for the fermionic theory as discussed in \cite{Aharony:2012nh, GurAri:2012is}. All parameters are defined in a dimensional reduction regularization scheme, used in \cite{Aharony:2012nh}. In terms of $\lambda = \frac{N}{k}$, the duality can be written as:
\begin{align}
\lambda_f  =   \lambda_b-\sign(\lambda_b), \\
N_f  =  N_b\frac{1-|\lambda|_b}{|\lambda|_b},
\end{align}
or
\begin{align}
\lambda_b  =  \lambda_f -\sign(\lambda_f),  \\
 N_b  =  N_f  \frac{1-|\lambda|_f}{|\lambda|_f}.
 \label{duality}
\end{align}
From these results, we have the simple relation $|\lambda_b|  =  1-|\lambda_f|$ and $\sign(\lambda_b)=-\sign(\lambda_f)$. 

As both sides of the theory are exactly solvable, the simplest way to verify the duality is to calculate correlation functions on both sides, which we illustrate below. 

The two-point function of the scalar primary,  which is defined as $J_f^{0} \equiv \bar{\psi}\psi$, in the fermionic theory is:
\begin{align}
\langle J_f^{0} (-q) J_f^{0} \rangle  =  -N_f \frac{\tan (\pi \lambda_f/2)}{4\pi \lambda_f} |q|.
\end{align}
In the critical bosonic theory, the two-point function is:
\begin{align}
\langle J_b^{0} (-q) J_b^{0} \rangle  =  -N_b \frac{4\pi \lambda_b}{\tan (\pi \lambda_b/2)} |q|.
\end{align}
These two-point functions determine the relative normalisation of the scalar operator in the two descriptions. We see that, taking $J_b^{0}=4 \pi \lambda_b J_f^{0}$\footnote{This relation makes sense because $\lambda$ and $J_f^0$ are odd under parity, while $J_b^0$ is even.}, the two point functions are identical:
\begin{align*}
 \langle J_b^{0} (-q) J_b^{0} \rangle =  -N_b \frac{4\pi \lambda_b}{\tan (\pi \lambda_b/2)} |q| \\
\langle J_f^{0} (-q) J_f^{0} \rangle  =  -N_b \frac{ \cot (\pi \lambda_b/2) }{4\pi \lambda_b} |q| \\
                                      =  N_f \sign(\lambda_b) \frac{ \tan (\pi \lambda_f/2) }{4\pi(1-|\lambda_b|)} |q| \\
                                      =  -N_f \frac{\tan (\pi \lambda_f/2)}{4 \pi \lambda_f} |q|
\end{align*}

The duality now implies that, for the three-point functions:
\begin{align}
\langle J_f^{0} (q_1) J_f^{0}(q_2) J_f^{0}(q_3) \rangle = \left(\frac{1}{4 \pi \lambda_b} \right)^3  \langle J_b^{0} (q_1) J_b^{0}(q_2) J_b^{0}(q_3) \rangle 
\end{align}

Using the results in \cite{Aharony:2012nh, GurAri:2012is}, we can explicitly compute that:
\begin{align*}
\langle J_f^{0} (q_1) J_f^{0}(q_2) J_f^{0}(q_3) \rangle  =  0 \\
\langle J_b^{0} (q_1) J_b^{0}(q_2) J_b^{0}(q_3) \rangle  =  0
\end{align*}
which agrees with the duality,.

There are additional predictions for three-point functions; for instance for the three-point function of vector and scalar operators, which are non-zero, and which can be tested on similar lines. 

Applying the duality to four-point functions, we obtain:
\begin{eqnarray}
\hspace{-3mm} \left(4 \pi \left(\lambda_f -\sign(\lambda_f) \right) \right)^4 \left \langle J_f^{0} (q_1) J_f^{0}(q_2) J_f^{0}(q_3) J_f^{0}(q_4) \right \rangle &=&  \left \langle J_b^{0} (q_1) J_b^{0}(q_2) J_b^{0}(q_3)  J_b^{0}(q_4) \right \rangle \label{duality-prediction}
\end{eqnarray}
In what follows, we directly calculate the LHS of \eqref{duality-prediction} (for a restricted class of external momenta) and obtain a finite answer in the limit $\lambda_f \rightarrow 1$ (when expressed in terms of $N_b$). The result can then be compared to a calculation the critical bosonic theory at $\lambda_b=0$ on the RHS and we find perfect agreement. 

As described below, and in \cite{Aharony:2012nh, GurAri:2012is}, the non-critical bosonic theory is dual to a critical fermionic theory. We also compare the critical fermionic theory to the non-critical bosonic theory and find agreement.

\label{sec:normalisation}
\section{The Exact Ladder Diagram}
\label{sec:ladder}
\begin{figure}
\centering % \begin{center}/\end{center} takes some additional vertical space
\includegraphics[scale=0.25]{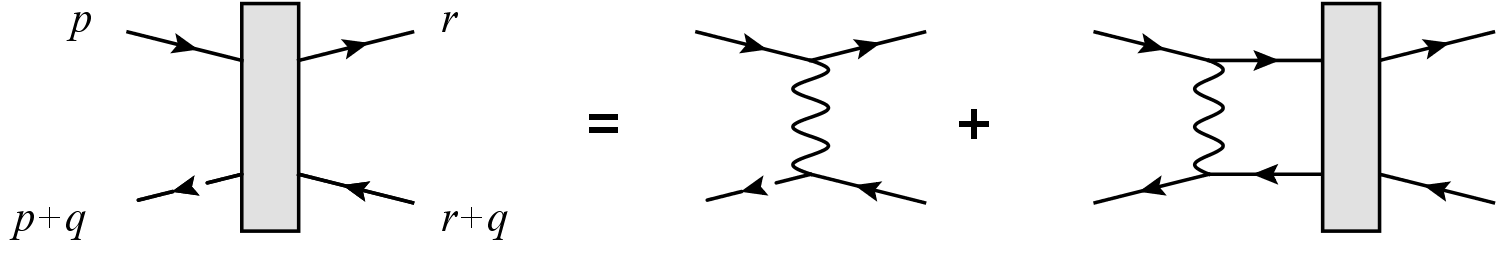}
\caption{\label{fig:i:1} The diagrammatic definition of the exact ladder diagram, which is the shaded box. All propagators in this diagram are exact propagators. In light cone gauge, this diagram is sufficient to calculate all planar correlation functions of single trace operators.}
\end{figure}

In this paper, we will exclusively use light-cone gauge\footnote{The conventions used in the following sections are those of \cite{Giombi:2011kc, Aharony:2012nh, GurAri:2012is}.   In particular, our light-cone gauge is defined in Euclidean space and described in detail in \cite{Giombi:2011kc}.  We also use the notation that $\gamma^{A}$ may be the $2\times2$ identity matrix $\mathbf 1$ or any of the three $\gamma^\mu$ and $p_s^2=p_1^2+p_2^2=2p_+p_-$.}. 
Two crucial features of light-cone gauge are that ghosts decouple and cubic vertices are absent. Therefore any planar correlation function can be evaluated to all orders in $\lambda$ given the exact propagator, first evaluated in \cite{Giombi:2011kc}, and the "exact ladder diagram" defined diagrammatically in Figure \ref{fig:i:1}, which we denote  by $\Gamma_{AB}(p,q,r) \gamma^A \otimes \gamma^B$. 

More precisely, we define $\Gamma_{AB}(p,q,r) \gamma^A \otimes \gamma^B$ as the following four-fermion interaction term in the quantum effective action for fermions, obtained after integrating out the gauge field in light cone gauge. 
\begin{equation}
S_{eff} = -\frac{1}{2} \int \frac{d^3p d^3q d^3r}{(2\pi)^9} \Gamma_{AB}(p,q,r) \bar{\psi}^i(-p) \gamma^A \psi_j(r) \bar{\psi}^j(-r-q) \gamma^B \psi_i(p+q).
\end{equation}

$S(p)$, the exact large $N$ fermion propagator, valid to all orders in $\lambda$, is defined via $\langle \psi_i(p) \bar{\psi}^j(q) \rangle = \delta_i^j (2\pi)^3 \delta^3(p-q) S(p)$ and is given by  \cite{Giombi:2011kc}:
\begin{equation}
S(p)= \left({-i \gamma^\mu p_\mu + i \lambda^2 \gamma^+ p^- + \lambda p_s \over p^2}\right).
\end{equation}
The gauge propagator is defined via $\langle A_\mu(p) A_\nu(-q) \rangle = \delta^{ab} (2\pi)^3 \delta^3(p-q)  G_{\nu\mu}(p) $ and is given by 
\begin{equation}
G_{+3}=\frac{4\pi i}{k} \frac{1}{p^+}.
\end{equation}
The self-consistent Schwinger-Dyson equation, given diagrammatically in Figure \ref{fig:i:1}, where all propagators are taken to be exact propagators $S(p)$, is:
\begin{eqnarray}
 && \Gamma_{AB}(p,q,r) \gamma^A \otimes \gamma^B  =  
\left(\frac{-2\pi i}{k} \right) \frac{1}{(p-r)^+} \left (\gamma^+ \otimes \gamma^3 - \gamma^3 \otimes \gamma^+ \right) \nonumber \\ &&
+ \left(\frac{-2N\pi i}{k} \right) \int \frac{d^3l}{(2\pi)^3} \frac{1}{l^+} \Gamma_{CD}(l,q,r) \nonumber \\ && \times  \left( \gamma^+ S(l) \gamma^C \otimes \gamma^D S(l+q) \gamma^3 - \gamma^3 S(l) \gamma^C \otimes \gamma^D S(l+q) \gamma^+ \right).
\end{eqnarray}

In terms of $S(p) = S_A(p) \gamma^A$, we have:
\begin{eqnarray}
 && \Gamma_{AB}(p,q,r) \gamma^A \otimes \gamma^B  =  
\left(\frac{-2\pi i}{k} \right) \frac{1}{(p-r)^+} \left (\gamma^+ \otimes \gamma^3 - \gamma^3 \otimes \gamma^+ \right) \nonumber \\ 
&&
- \left(2\lambda\pi i \right) \int \frac{d^3l}{(2\pi)^3} \frac{1}{l^+} \Gamma_{CD}(l,q,r)S_E(l) S_F(l+q) \nonumber \\ 
&& \times  \left( 
\gamma^+  \gamma^F \gamma^C 
\otimes 
\gamma^D \gamma^E \gamma^3 
- 
\gamma^3 \gamma^F \gamma^C 
\otimes 
\gamma^D  \gamma^E \gamma^+ 
\right) \label{ladder}
\end{eqnarray}

% Note that $\Gamma_{AB}(p,q,r) = \Gamma_{BA}(q,p,-r)$.

\subsection{Rewriting the Schwinger-Dyson equation}
\label{rewrite}
There are 16 components of $\Gamma_{AB}$, which appear to be coupled. We now obtain an alternate expression for $\Gamma_{AB}$ that ``diagonalizes'' the Schwinger-Dyson equation \eqref{ladder} and shows that most of the 16 components are not independent.

For this purpose it is convenient to define:
\begin{equation}
4\pi i A_P{}^Q(p,q,r)\gamma^P = \Gamma_{AB}(p,q,r)\gamma^A \gamma^Q \gamma^B .
\label{Adef}
\end{equation}
It is easy to see that the inverse relation is 
\begin{equation}
4\pi i \text{ Tr} \left(\gamma_{A} \gamma^P \gamma_{B} \gamma_Q\right) A_P{}^Q(p,q,r) = \Gamma_{AB}(p,q,r) 
\label{Adef3}
\end{equation}
(where `Tr' denotes a trace over the gamma matrices.)

Let us also define, following \cite{Giombi:2011kc} 
\begin{equation}
\label{contraction}
 H_+(Y) =  \gamma^3 Y \gamma^+ -\gamma^+ Y \gamma^3= 2( Y_I \gamma^+ - Y_- I).
\end{equation}
The Schwinger-Dyson equation \eqref{ladder} can be re-written as
\begin{eqnarray}
4\pi i A_P{}^Q(p,q,r) \gamma^P
 & = & \Gamma_{AB}(p,q,r) \gamma^A \gamma^Q \gamma^B \\
& = &  
\left(\frac{-2\pi i}{k} \right) \frac{1}{(p-r)^+} \left (\gamma^+ \gamma^Q \gamma^3 - \gamma^3 \gamma^Q \gamma^+ \right) \nonumber 
\\ 
&&
+ \left(\frac{-2N\pi i}{k} \right) \int \frac{d^3l}{(2\pi)^3} \frac{l}{l^+} \Gamma_{CD}(l,q,r)S_E(l) S_F(l+q) \nonumber 
\\ 
&& \times  \left( 
\gamma^+  \gamma^F \gamma^C 
\gamma^Q 
\gamma^D \gamma^E \gamma^3 
- 
\gamma^3 \gamma^F \gamma^C 
\gamma^Q
\gamma^D  \gamma^E \gamma^+ 
\right)  \\
& = & 
\left(\frac{2\pi i}{k} \right) \frac{1}{(p-r)^+} H_+(\gamma^Q) \nonumber 
\\ 
&&
+ \left(\frac{2N\pi i}{k} \right) \int \frac{d^3l}{(2\pi)^3} \frac{1}{l^+} \Gamma_{CD}(l,q,r)S_E(l) S_F(l+q) \nonumber \times  \nonumber \\ &&
H_+(  \gamma^F \gamma^C 
\gamma^Q 
\gamma^D \gamma^E )
 \label{boot3} 
\\
 A_P{}^Q(p,q,r) \gamma^P
& = & 
\left(\frac{1}{2k} \right) \frac{1}{(p-r)^+} H_+(\gamma^Q) \nonumber 
\\ 
&&
+  2\pi i \lambda   \int \frac{d^3l}{(2\pi)^3} \frac{1}{l^+} A_P{}^Q(l,q,r)S_E(l) S_F(l+q) \nonumber 
 \times  \nonumber \\ &&
H_+(  \gamma^F  
\gamma^P 
 \gamma^E )
 \label{boot4}
\end{eqnarray}
Because $H_+$ contains only the identity and $\gamma^+$ components,this means that 
\begin{equation}
 A_3{}^Q(p,q,r) = A_-{}^Q(p,q,r) = 0
\end{equation}
for all $Q$. This translates into $8$ linear equations relating various of the 16 components of $\Gamma_{AB}$.   Moreover, the 8 $A_+{}^Q$ and $A_I{}^Q$ are the only non-vanishing components of $A$, and they are independent for different values of $Q$. It is also consistent to set $A_P{}^3=0$ and $A_P{}^+=0$ so we have only 4 equations, which are 2 pairs of 2 coupled integral equations.

Evaluating \eqref{boot4} explicitly, we obtain:
\begin{eqnarray}
 A_+{}^Q(p,q,r) \gamma^+ + A_I{}^Q(p,q,r) 
&=&
\left(\frac{1}{2k} \right) \frac{1}{(p-r)^+} H_+(\gamma^Q)
\nonumber 
\\ 
&& \hspace{-2cm}
+ 2\pi i \lambda  \int \frac{d^3l}{(2\pi)^3} \frac{1}{l^+} A_+{}^Q(l,q,r)S_E(l)S_F(l+q) \nonumber 
 \times  
H_+(  \gamma^F  
\gamma^+ 
 \gamma^E )
\nonumber 
\\ 
&& \hspace{-2cm}
+ 2\pi i \lambda  \int \frac{d^3l}{(2\pi)^3} \frac{1}{l^+} A_I{}^Q(l,q,r)S_E(l)S_F(l+q)  
 \times  
H_+( \gamma^F  
 \gamma^E ) \label{ieqn}
\end{eqnarray}

\subsection{Evaluating the exact ladder diagram when $q_\pm=0$}

We have not yet been able to solve this integral equation for arbitrary $q$. However, if we restrict ourselves to $q_\pm=0$ it is possible to obtain a solution, which will enable us to calculate the four-point function of scalar primaries for a restricted class external momenta.

\begin{figure}[h]
\centering % \begin{center}/\end{center} takes some additional vertical space
\includegraphics[scale=0.15]{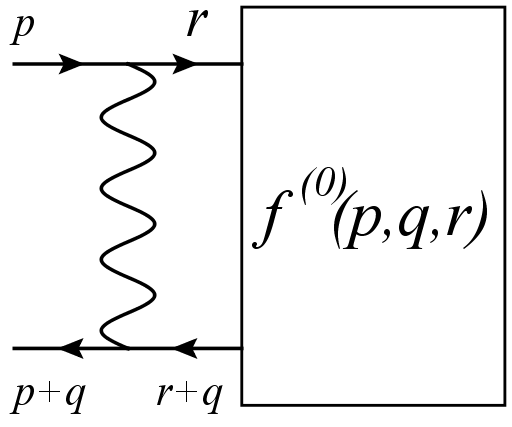}
\caption{\label{fig:i:2} A contraction of the tree-level ladder diagram corresponding to   Equation \eqref{contr}. Note that there is no integration over momenta.}
\end{figure}

To motivate our ansatz for the solution, we note that the results of section \ref{rewrite} can also be thought of diagrammatically as follows: Let $f^{(0)}(p,q,r)$ be any $2\times2$ matrix (with spinor indices) that is a function of $p$, $q$ and $r$. We think of $f^{(0)} \delta^i_j$ as representing an arbitrary "contraction" of the ladder diagram on the right, so that the tree level ladder diagram acting on $f^0 \delta^l_m$ is given by 
\begin{equation}
\frac{N}{2} \gamma^\nu f^{(0)}(p,q,r) \gamma^\mu G_{\mu\nu}(p-r)\delta^i_j =\frac{N}{2}\delta^i_j G_{+3}(p-r)H_+\left(f^{(0)}(p,q,r)\right) \label{contr}
\end{equation}
as pictured in Figure \ref{fig:i:2}. 

\begin{figure}[h]
\centering % \begin{center}/\end{center} takes some additional vertical space
\includegraphics[scale=0.15]{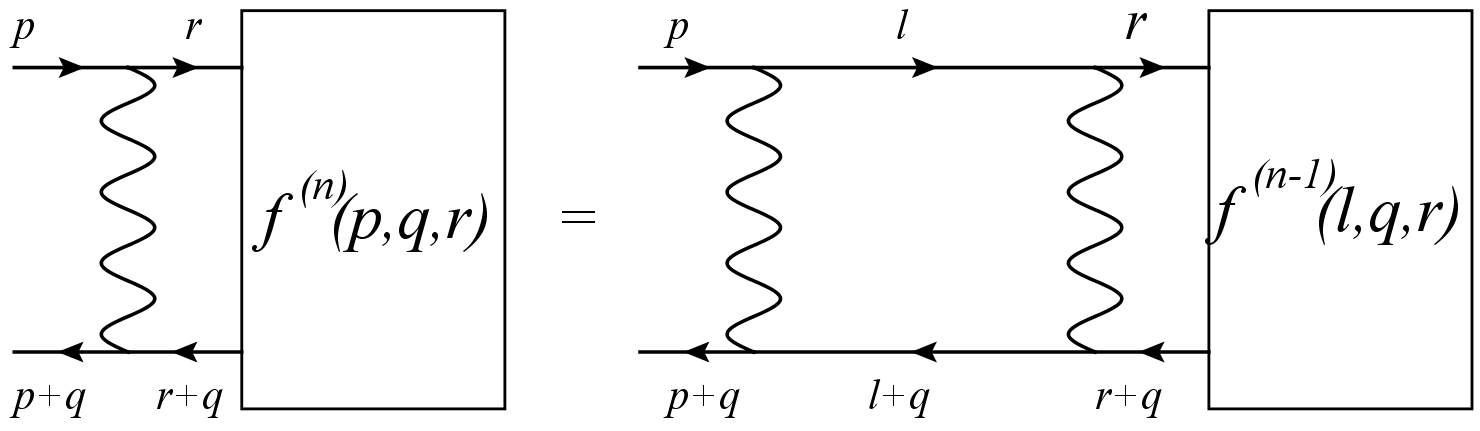}
\caption{\label{fig:i:3} The diagrammatic relation for $f^{(n)}$ in terms of $f^{(n-1)}$.}
\end{figure}

We then define $f^{(n)}(p,q,r)$ (which can be thought of as the ladder diagram with $n$ "rungs", contracted with $f^{(0)}$ on the right) recursively in terms of $f^{(n-1)}$: 
\begin{equation}
\gamma^\nu f^{(n)}(p,q,r) \gamma^\mu G_{\mu\nu}(p-r)=  \frac{N}{2} \gamma^\nu \left(\int \frac{d^3l}{(2\pi)^3}  S(l)  \gamma^\sigma f^{(n-1)}(l,q,r)\gamma^\rho G_{\rho \sigma}(l-r) S(l+q) G_{\mu\nu}(p-l) \right) \gamma^\mu  \label{pre-ladder}
\end{equation}
as pictured in Figure \ref{fig:i:3}. 

Because $G_{+3}=-G_{3+}$ are the only nonzero components of $G_{\mu\nu}$, only two components of $f^{(n-1)}$ contribute to $f^{(n)}$, which are
\begin{equation}
 H_+\left(f^{(n)} \right)=2f_I^{(n)} \gamma^+ - 2f_-^{(n)} \mathbb{I}.
\end{equation}
In terms of these variables, the equation \eqref{pre-ladder} is:
\begin{equation}
{H_+(f^{(n)}(p,q,r)) \over (p-r)^+ } 
=  2\pi i \lambda\int \frac{d^3l}{(2\pi)^3}   H_+ \left( S(l)  H_+\left(f^{(n-1)}(l,q,r)\right)S(l+q) \right)\frac{1}{(l-r)^+ (p-l)^+}     \label{pre-ladder2}
\end{equation}

The infinite sum $\sum f^{(n)}$ is related to $A_P^Q$ defined in \eqref{Adef} of the previous subsection and $f_I^{(0)}$  and $f_-^{(0)}$ as follows:
\begin{equation}
\frac{\sum f^{(n)}_I}{k(p-r)^+}=A_+^If_I^{(0)}+A_+^-f_-^{(0)}, ~~~
 -\frac{\sum f^{(n)}_-}{k(p-r)^+}=A_I^I  f_I^{(0)}+A_I^-f_-^{(0)}. 
 \label{eq:19}
 \end{equation}. 
 
Let us choose $f^{(0)}$ (which is an arbitrary matrix) such that $f_-^{(0)}= - c f_I^{(0)} r^+$.  The parameter $c$ is arbitrary, and introduced for convenience: when we set $c=0$, equation \eqref{eq:19} determines $A_+^I$ and $A_I^I$ and when $c \rightarrow +\infty$, equation \eqref{eq:19} determines $A_+^-$ and $A_I^-$.

We note that it is consistent to assume $f^{(n)}(p,q,r)$ is independent of $p_3$. Evaluating equation \eqref{pre-ladder}, including the $l_3$ integral, we have:
 \begin{equation}
 \frac{f_I^n(p)}{(p-r)^+}=-2i \lambda  \int \frac{d^2l}{(2\pi )^2}\frac{f_I^{n-1}(l)\left(q_3+2i \lambda  l_s\right) l^+ - f_-^{n-1}(l)\left(2\left(1-\lambda ^2\right)l_s^2\right)}{\left(q_3^2+4l_s^2\right)l_s}\frac{1}{(p-l)^+(l-r)^+}
 \end{equation}
 \begin{equation}
 \hspace{-8 mm}
 \frac{f_-^n(p)}{(p-r)^+}=-2i \lambda  \int \frac{d^2l}{(2\pi )^2}\frac{f_I^{n-1}(l)2l^{+2}-f_-^{n-1}(l)\left(2i \lambda  l_s-q_3\right)l^+}{\left(q_3^2+4l_s^2\right)l_s}\frac{1}{(p-l)^+(l-r)^+}
 \end{equation}
After integrating to obtain the first few terms, we find $f_I^{(n)}(p)$ and $f_-^{(n)}(p)$ to be of the form:
 \begin{align}
 f_I^{(n)}(p)=a_1^{(n)}+a_2^{(n)} \frac{r^+}{p^+}  \label{eq:24a}    \\
 f_-^{(n)}(p)=-b_1^{(n)}p^+-b_2^{(n)} r^+ 
 \label{eq:24}
 \end{align}
 
We can sum the series to obtain:
\begin{equation} 
\sum f_I^{(n)}(p)- f_I^{(0)} =
-2i \lambda  \int \frac{d^2l}{(2\pi )^2}\frac{\left(\sum f_I^n(l)\left(q_3+2i \lambda  l_s\right)l^+ -\sum f_-^n(l)\left(2\left(1-\lambda ^2\right)l_s^2\right)\right)}{\left(q_3^2+4l_s^2\right)l_s}\frac{(p-r)^+}{(p-l)^+(l-r)^+}
\label{eq:25}
\end{equation}
\begin{equation} 
\sum f_-^{(n)}(p)- f_-^{(0)}=-2i \lambda  \int \frac{d^2l}{(2\pi )^2}\frac{\left(\sum f_I^{(n)}(l)2l^{+2}-\sum f_-^{(n)}(l)\left(2i \lambda  l_s-q_3\right)l^+\right)}{\left(q_3^2+4l_s^2\right)l_s}\frac{(p-r)^+}{(p-l)^+(l-r)^+}
\label{eq:26}
 \end{equation}
  From \eqref{eq:24a} and \eqref{eq:24}  we make the  ansatz
\begin{equation}
  \sum f_I^{(n)}(p)=\left(a_1+a_2\frac{r^+}{p^+}\right)f_I^{(0)}  \hspace{10 mm}       \sum f_-^{(n)}(p)=-\left(b_1p^++b_2r^+\right)f_I^{(0)}.
  \end{equation}
We now use the identity
   \begin{equation}
 {\frac{1}{(p-l)^+(l-r)^+}} ={\frac{1}{(p-r)^+}\left(\frac{1}{(p-l)^+}+ \frac{1}{(l-r)^+}\right)}
  \end{equation}
and, following \cite{Aharony:2012nh}, introduce the dimensionless variables:
\begin{align} 
 y=\frac{2p_s}{|q_3|} \hspace{5 mm}
 x=\frac{2r_s}{|q_3|} \hspace{5 mm}
 t=\frac{2l_s}{|q_3|} \hspace{5 mm}
 \hat{\Lambda}=\frac{2\Lambda_s}{|q_3|} \hspace{5 mm}
 \hat{\lambda}=\lambda \text{ sign}(q_3)
 \label{eq:27}
\end{align} 
 to carry out the angular integrals and rewrite \eqref{eq:25} and \eqref{eq:26}  as
 \begin{align}
  \left(\left(a_1+a_2\frac{r^+}{p^+}\right)-1\right) &=& \int_{x}^{y} \left(a_1\left(1+i \hat{\lambda } t \right)+\frac{b_1}{2}\left(1-\hat{\lambda}^2\right)t^2\right) \, dt+ \int_{x}^0
\left(a_2(1+i \lambda  t)+b_2\right) \, dt\nonumber \\ && +
  \frac{r^+}{p^+}\int_0^{y} \left(a_2\left(1+i \hat{\lambda  }t\right)+\frac{b_2}{2}\left(1-\hat{\lambda }^2\right)t^2\right) \, dt 
  \label{eq:30}
 \end{align}
 \begin{align}
 \left(b_1p^++b_2r^+\right)-c r^+ =&&  
 r^+\left(\int _{x}^{\Lambda }\left(2a_1+b_1\left(i \hat{\lambda } t-1\right)\right)dt \right.
  \left.+\int_{x}^{y} \left(2a_2+b_2\left(i \hat{\lambda } t-1\right)\right) \, dt \right)\nonumber \\ &&
  + p^+\int_{\Lambda }^{y} \left(2 a_1+b_1\left(i \hat{\lambda } t -1\right)\right) \, dt. 
  \label{eq:31}
 \end{align}
 
To solve these coupled four-variable equations (which could in principle have been obtained directly from equation \eqref{ieqn} with the approporiate ansatz for $A_P^Q$) we differentiate and the resulting differential equation decouples into two sets of two variable coupled equations.

Equating coefficients of $\frac{r^+}{p^+}, p^+$ and $r^+$, we obtain:
\begin{align} 
 && \frac{\partial a_1}{\partial y}=-\left(a_1\left(1+i \hat{\lambda } y\right)+\frac{b_1}{2}\left(1-\hat{\lambda }^2\right)y^2\right)
\nonumber \\ &&
 \frac{1}{2}\frac{\partial b_1}{\partial y}=\left(a_1+\frac{b_1}{2}\left(i \hat{\lambda } y-1\right)\right)
 \nonumber \\ &&
 \frac{\partial a_2}{\partial y}=-\left(a_2\left(1+i \hat{\lambda } y\right)+b_2\left(1-\hat{\lambda }^2\right)y^2\right)
 \nonumber \\ &&
 \frac{1}{2}\frac{\partial b_2}{\partial y}=\left(a_2+b_2\left(i \hat{\lambda } y-1\right)\right)
 \end{align}
The general solution of the set is
\begin{eqnarray}
a_i &=& \alpha _i\left(\beta _i(1-i \hat{\lambda}  y )-(1+i \hat{\lambda}  y)e^{-2i \hat{\lambda} \arctan(y)}\right)
\nonumber \\ 
 b_i&=&2\alpha_i\left(\beta _i+ e^{-2i \hat{\lambda} \arctan(y)}\right)
 \end{eqnarray}
 where the subscript $i=1,2$.
 Requiring that the solutions satisfy the integral equations \eqref{eq:30} and \eqref{eq:31} fixes the integration constants:
\begin{eqnarray} 
 \beta_1 &=&-e^{-2i \hat{\lambda}\arctan[\hat{\Lambda}]} \nonumber \\  \beta_2 & = & 1
\nonumber \\ 
 \alpha _1 & = & \frac{\frac{c}{2}\left((1-i \hat{\lambda} x)-(1+i \hat{\lambda} x)e^{-2i \hat{\lambda}   \arctan[x]}\right)-\left(1+e^{-2i \hat{\lambda} \arctan[x]}\right)}{2e^{-2i \hat{\lambda}   \arctan[x]}\left(1+e^{-2i \hat{\lambda}   \arctan[\hat{\Lambda}]}\right)}
\nonumber \\ 
\alpha_2 &=& 
\frac{\frac{c}{2}\left((1-i \hat{\lambda}  x)e^{-2i \hat{\lambda} \arctan[\hat{\Lambda}]}+(1+i \hat{\lambda}  x)e^{-2i \hat{\lambda} \arctan [x]}\right)
-\left(e^{-2i \hat{\lambda} \arctan[\hat{\Lambda}]}-e^{-2i \hat{\lambda} \arctan[x]}\right)}{2e^{-2i \hat{\lambda} \arctan[x]}\left(1+e^{-2i\hat{\lambda} \arctan(\hat{\Lambda})}\right)}.
\label{coeffc}
\end{eqnarray}

To determine $A_+^I$ and $A_I^I$ we set $c=0$ and to determine $A_+^-$ and $A_I^-$ we take the limit $c\rightarrow \infty$ (i.e., equate coefficients of $c$ on both sides of Equation \eqref{eq:19}). Writing the answers in the form:
$$\tilde{A}_C^B =2k e^{-2i \hat{\lambda} \arctan[x]}\left(1+e^{-2i\hat{\lambda} \arctan(\hat{\Lambda})}\right)(p-r)^+ A_C^B  $$
we have:
  \begin{eqnarray}
  \tilde{A}_I^I &=&\left(e^{-2i \hat{\lambda}   \arctan[\hat{\Lambda}]}-e^{-2i \hat{\lambda}   \arctan[y]}\right)\left(1+e^{-2i \hat{\lambda}   \arctan[x]}\right) \left(\frac{2p^+}{q_3}\right)
 \nonumber \\ && +
 \left(e^{-2i \hat{\lambda}   \arctan[x]}-e^{-2i \hat{\lambda}   \arctan[\hat{\Lambda}]}\right)\left(1+e^{-2i \hat{\lambda}   \arctan[y]}\right)\left(\frac{2r^+}{q_3}\right)
 \label{ladder1} \\
\tilde{A}_I^- &=& -\left(e^{-2i \hat{\lambda}   \arctan[\hat{\Lambda}]}-e^{-2i \hat{\lambda}   \arctan[y]}\right)\left(( i\hat{\lambda} x-1)+(1+ i\hat{\lambda} x)e^{-2i \hat{\lambda}   \arctan[x]}
\right)
\left(\frac{p^+}{r^+}\right)
\nonumber \\ && -
 \left((1+ i\hat{\lambda} x )e^{-2i \hat{\lambda}   \arctan[x]}-( i\hat{\lambda} x -1)e^{-2i \hat{\lambda}   \arctan[\hat{\Lambda}]}\right)\left(1+e^{-2i \hat{\lambda}  \arctan[y]}\right)
 \label{ladder2} \\
\tilde{A}_+^I &= & \left((1+ i\hat{\lambda} y )e^{-2i \hat{\lambda}   \arctan[y]}-( i\hat{\lambda} y-1)e^{-2i \hat{\lambda}  \arctan[\hat{\Lambda}]}\right)\left(1+e^{-2i
\hat{\lambda}   \arctan[x]}\right)
\nonumber \\ && +
\left(e^{-2i \hat{\lambda}   \arctan[\hat{\Lambda}]}-e^{-2i \hat{\lambda}   \arctan[x]}\right)\left(( i\hat{\lambda} y-1)+(1+ i\hat{\lambda} y)e^{-2i \hat{\lambda}   \arctan[y]}\right)\left(\frac{r^+}{p^+}\right)
\label{ladder3}\\
 \hspace{-3 mm}
\tilde{A}_+^- &=& -\left((1+ i\hat{\lambda} y )e^{-2i \hat{\lambda}   \arctan[y]}-( i\hat{\lambda} y-1)e^{-2i \hat{\lambda}   \arctan[\hat{\Lambda}] ]}\right)
 \nonumber \\ && 
 \times \left(\left((1+ i\hat{\lambda} x )e^{-2i \hat{\lambda}   \arctan[x]}+( i\hat{\lambda} x-1)\right)
\left(\frac{q_3}{2r^+}\right)\right)
 \nonumber \\ &&
 -
\left(( i\hat{\lambda} x-1)e^{-2i \hat{\lambda} \arctan[\hat{\Lambda}]}-(1+ i\hat{\lambda} x )e^{-2i \hat{\lambda}   \arctan[x]}\right)
  \nonumber \\ &&
\times \left(( i\hat{\lambda} y-1)+(1+ i\hat{\lambda} y )e^{-2i \hat{\lambda}   \arctan[y]}\right)\left(\frac{q_3}{2p^+}\right)
\label{ladder4}
\end{eqnarray}

In Appendix A, we evaluate the exact vertex for the scalar primary using this result.

\section{Four-Point Correlation Functions} 

We now proceed to calculate the (gauge-invariant and parity-invariant) four-point function of the scalar primary. In this section, we evaluate the four-point function in the free fermionic theory, the interacting fermionic theory and the critical interacting fermionic theory. In section \ref{bosonic_theory_section}, we then evaluate the four-point function in the free and critical bosonic theories to test the duality. The four-point functions depend on external momenta $q^{(i)}$; as discussed in the previous section, our calculations are only valid in the special case of $q_\pm=0$ (i.e., only $q_3 \neq 0$). Hence, in what follows, we drop all spacetime-indices and label the four external momenta as $q_1$, $q_2$, $q_3$ and $q_4$.

\subsection{Free Fermionic Theory}

%\begin{figure}
%\centering % \begin{center}/\end{center} takes some additional vertical space
%\includegraphics[scale=0.75]{fourcorrelator1}
%\caption{\label{fig:i:4} Diagrams contributing to the four-point function in the free theory.}
%\end{figure} 

The free ($\lambda_f=0$) four-point function in the $U(N_f)$ fermionic theory is given by 
\begin{eqnarray}
 F(q_1,q_2,q_3,q_4) & = & \langle J^{(0)}(-q_1)J^{(0)}(-q_2)J^{(0)}(-q_3)J^{(0)}(-q_4) \rangle \nonumber \\ &=&-N_f \int \frac{d^3k}{(2\pi )^3}\text{ Tr} 
\left( S_0(k)S_0(k+q_1)S_0(k+q_1+q_2) S_0(k+q_1+q_2+q_3)\right)  \nonumber \\ && + \text{ permutations}
\end{eqnarray}
where $S_0(q)=(i\slashed q)^{-1}$ is the free fermion propagator and all external momenta are restricted in the 3 direction. 

Evaluating this, we obtain
\begin{equation}
F(q_1,q_2,q_3,q_4) =N_f\frac{1}{8(q_1+q_2)(q_1+q_4)}(\mid q_1\mid-\mid q_2 \mid+\mid q_3 \mid-\mid q_4 \mid )+\text{5 permutations}
\label{eq:28}
\end{equation}
which can be simplified to:
\begin{equation}
F(q_1,q_2,q_3,q_4) =N_f\frac{q_1|q_1|+q_2|q_2|+q_3|q_3|+q_4|q_4|}{2(q_1+q_2)(q_1+q_3)(q_2+q_3)} \delta(q_1+q_2+q_3+q_4)
\label{eq:41}
\end{equation}

\subsection{Interacting Fermionic Theory}
\label{Interacting}
We now proceed to calculate the four-point function in the interacting theory. 

\begin{figure}
\centering % \begin{center}/\end{center} takes some additional vertical space
\includegraphics[scale=0.7]{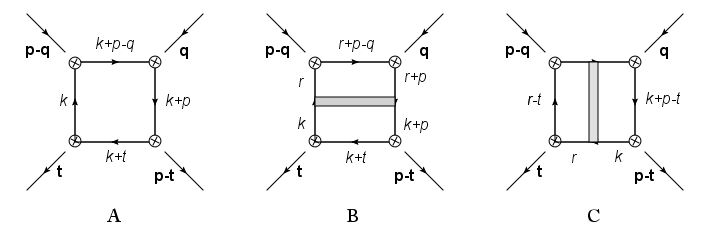}
\caption{\label{fig:i:5} Diagrams in the interacting theory.}
\end{figure}

In the interacting theory, our basic ingredients are the exact propagator, the exact $J^{(0)}$ vertex $V$ \cite{GurAri:2012is}, and the ladder diagram in section \ref{sec:ladder}. The correlator can be written as as a sum of Diagrams $A$, $B$ and $C$ in Figure \ref{fig:i:5} (where it is understood that all propagators are exact), 
\begin{equation}
\langle J^{(0)}(p-q)J^{(0)}(-t)J^{(0)}(t-p)J^{(0)}(q) \rangle = [\text{A}]+[\text{B}] + [\text{C}].
\end{equation}
The diagrams $B$ and $C$ involve the ladder diagram. 

Diagram $A$ is given by
\begin{align}
(\text{A}) = -N \int \frac{d^3k}{(2\pi)^3} \text{Tr}\left[V(p-q)S(k)V(-t)S(k+t)V(t-p)S(k+p)V(q)S(k+p-q)\right] \nonumber \\ + (5 \text{ permutations})
\label{eq:46}
\end{align}
and diagrams $B$ and $C$ are given by
\begin{align}
(\text{B}) & = & N \int \frac{d^3k}{(2\pi)^3} \text{Tr}[S(r+p)V(q) S(r-p+q)V(p-q) S(r)\gamma^\mu S(k)V(-t)S(k+t)V(t-p) \nonumber \\ && S(k+p)\gamma^\nu \Gamma_{\mu\nu}(k,p,r)] + (5 \text{ permutations}) \\
(\text{C}) & = & N \int \frac{d^3k}{(2\pi)^3} \text{Tr} [(S(r+p-q-t)V(p-q)S(r-t)V(-t)S(r)\gamma^\mu S(k)V(t-p)  S(k+p-t) \nonumber \\ && V(q) S(k+p-q-t)\gamma^\nu \Gamma_{\mu\nu}(k,p-q-t,r)] + (5 \text{ permutations})
\label{eq:47}
\end{align}
It is difficult to solve this integral in closed from for arbitrary $p$, $q$, and $t$. To obtain humanly readable answers that can be easily compared to the bosonic theory, we observe that the limit where two momenta ($ q\rightarrow  0^+ $ and $t\rightarrow  0^+ $ ) vanish is relatively tractable.

Let us first consider the diagrams where the two non-vanishing external momenta are ``diagonal" (as depicted in Figure \ref{fig:i:5} when $ q\rightarrow  0^+ $ and $t\rightarrow  0^+ $). In this limit, the integral is solvable. We find the 2 ``diagonal" permutations of diagram $A$ are given by
\begin{align}
 && -N \int \frac{d^3k}{(2\pi)^3} \text{Tr} [V(p-q)S(k)V(-t)S(k+t) V(t-p)S(k+p)V(q)S(k+q)] \nonumber
 \\&& =  -N \frac{\sec(\frac{\pi \lambda}{2})^2 (\pi \lambda + \sin(\pi \lambda))}{4( 2\pi \lambda |p|)} 
 \label{eq:29}
 \end{align}
For diagrams $B$ and $C$, we use \eqref{eq:19}and \eqref{ladder1}-\eqref{ladder2} for the ladder expressions, and obtain
\begin{equation}
\text{2 permutations of B} =N \sec^3\left(\frac{\pi\lambda}{2}\right)\frac{(2 \pi\lambda \cos\left(\frac{\pi\lambda}{2}\right)-7 \sin \left(\frac{\pi\lambda}{2}\right) + \sin \left(\frac{3\pi\lambda}{2}\right))}
{32p \pi \lambda}
\label{eq:59}
\end{equation}
Adding \eqref{eq:29} and \eqref{eq:59} for B as well as C gives
\begin{equation}
-N\frac{2 \sec^2\left(\frac{\pi \lambda}{2}\right) \tan\left(\frac{\pi \lambda}{2}\right) }{4|p| \pi \lambda}
\label{eq:44}
\end{equation}

The next step is to evaluate the remaining permutations of diagrams $A$, $B$ and $C$ with two adjacent non-vanishing external momenta (i.e., permutations of the external momenta not depicted in Figure $4$). These integrals are more nontrivial, but can be obtained by using the substitution $q \rightarrow p-q$ in \eqref{eq:46} and \eqref{eq:47} which evaluates to 
\begin{equation}
-N \frac{1}{|p|}\left(\frac{\sec^2\left(\frac{\pi \lambda}{2}\right) \tan\left(\frac{\pi \lambda}{2}\right) }{4 \pi \lambda}+h(\lambda) \right)
\end{equation}
where $h(\lambda)$ is 
\begin{align}
h(\lambda) = \frac{i}{4\left(1+e^{i \pi \lambda}\right)^2} 
     \Big(1 + 2 e^{i \pi \lambda} \pi \lambda \cot \left(\pi \lambda \right) + \lambda \psi \left(\frac{1-\lambda}{2}\right)-\lambda \psi \left(\frac{-\lambda}{2}\right) + \nonumber \\
   e^{2i \pi \lambda} \left( 1 + \lambda \psi \left(\frac{\lambda}{2}\right) - \lambda \psi \left(\frac{1+\lambda}{2}\right)\right) \Big)
  \label{hresult}
\end{align}
and $\psi(x)=\frac{\Gamma'(x)}{\Gamma(x)}$ is the Digamma function.
It can be seen that the above equation goes to 0 for $\lambda \rightarrow 0$ and decreases approximately as $-\tan^2(\frac{\pi \lambda}{2})$. This property will be later be of use in the critical theory.

In summary, after adding all the diagrams we get the following result
\begin{equation}
\langle J^{(0)}(p-q)J^{(0)}(-t)J^{(0)}(t-p)J^{(0)}(q) \rangle=-N\frac{2 \sec^2\left(\frac{\pi \lambda}{2}\right) \tan\left(\frac{\pi \lambda}{2}\right) }{|p| \pi \lambda}-\frac{4 N h(\lambda)}{|p|}
\label{eq:45}
\end{equation}
\begin{figure}
\centering % \begin{center}/\end{center} takes some additional vertical space

\includegraphics[scale=0.8]{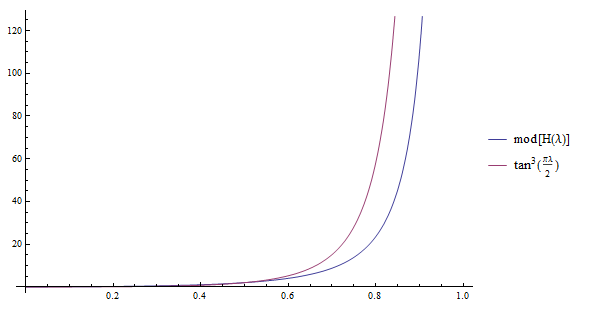} 

\label{fig:fermion graphs}
\caption{ The plot of modulus of $H(p,\lambda)$ function and $\tan^3[\frac{\pi \lambda}{2}]$ vs $\lambda$ for momenta $p=1$. It is clear that the magnitude of $H(\lambda)$ rises slower than $ \tan^3[\frac{\pi \lambda}{2}] $.}
\end{figure}

It is worthwhile to notice that both terms in the R.H.S of \eqref{eq:45} are parity invariant.
We can draw a parallel between \eqref{eq:28} ,\eqref{eq:41} and  \eqref{eq:44} ,\eqref{eq:45} respectively.

It is natural to conjecture that this expression generalises to the following expression for general momenta:
\begin{align}
\langle J^{(0)}(-q_1)J^{(0)}(-q_2)J^{(0)}(-q_3)J^{(0)}(-q_4) \rangle=\frac{2\sec^2\left(\frac{\pi \lambda}{2}\right) \tan\left(\frac{\pi \lambda}{2}\right) }{ \pi \lambda}F(q_1,q_2,q_3,q_4)\nonumber \\ + H(q_1,q_2,q_3,q_4,\lambda)
\label{main_result}
\end{align}
where $F(q_1,q_2,q_3,q_4)$ is the four-point function of the scalar operator in the free fermionic theory and $H(q_1,q_2,q_3,q_4,\lambda)$ is an additional structure, which goes to zero as $\lambda \rightarrow 0$.

\subsection{Critical Fermionic Theory}
%CRITICAL FERMIONIC THEORY
We now consider the four point function of the scalar primary in the critical fermionic theory described in \cite{Aharony:2012nh}, \cite{GurAri:2012is}, which is conjectured to be dual to the (non-critical) bosonic theory.

Let us briefly review the definition of the critical fermonic theory, which at zero-coupling is essentially the Gross-Neveu model (in three-dimensions): We introduce a field $\sigma$ (without a kinetic term) that couples to the scalar primary as $S_\sigma = \int d^3x \sigma \bar{\psi}\psi$ and perform a path integral over $\sigma$. The equation of motion for $\sigma$ is $\bar{\psi}\psi = 0$; therefore, instead of $\bar{\psi}\psi$, the single trace scalar primary operator in the critical theory is $\sigma$. Notice that, $\bar{\psi}\psi$ has scaling dimension 2, so $\sigma$  has scaling dimension $1$, which matches the scaling dimension of the scalar primary $J^{(0)}_b=\bar{\phi}\phi$ in the non-critical bosonic theory, as required for a duality. Because $\sigma$ has mass dimension $1$, there is the possibility of an additional marginal coupling of the form $\int d^3x N\frac{\lambda^F_6}{3!} \sigma^3$, which is related to the marginal $\phi^6$ coupling in the bosonic theory. 

In the large $N$ limit, the exact two point function of $\sigma$ is clearly related the inverse of the two point function of $\bar{\psi}\psi$, via 
\begin{equation}
G(q)=\langle \sigma(-q)\sigma \rangle = \left(-\langle J^{(0)}_f(-q) J^{(0)}_f \rangle_{\text{non-critical}} \right)^{-1}, \label{a}
\end{equation}
where two-point function of the scalar primary $J_f^{(0)}$ in the non-critical theory is
\begin{equation}
\langle J_f^{(0)}(-q)J_f^{(0)}\rangle=-\frac{N_f\tan \left(\frac{\pi \lambda_f}{2} \right)}{4\pi \lambda_f} |q|.
\end{equation}

\begin{figure}[h]\centering % \begin{center}/\end{center} takes some additional vertical space
\begin{tabular}{ccc}
\includegraphics[scale=0.4]{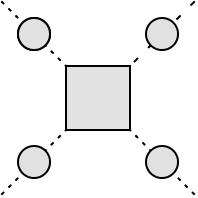} &  
\includegraphics[scale=0.4]{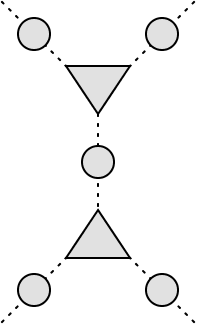} & 
\includegraphics[scale=0.4]{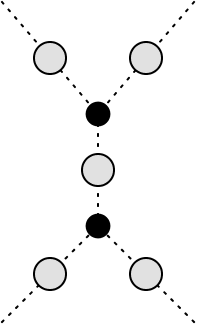}

\end{tabular}

\caption{ The diagrams that contribute to the four-point function of $\sigma$  in the critical fermionic theory. The shaded square is the exact scalar four-point function in the non-critical theory, the shaded triangle is the exact three-point function in the non-critical theory and the dashed line with a shaded circle is the $\langle \sigma\sigma \rangle$ propagator. The last diagram includes the contribution from the marginal $\lambda_6$ coupling, represented by black dot. (There is also another diagram with only one $\lambda_6$ vertex and one exact three point function that is not pictured.) \label{fig:critical_fermions}
}
\end{figure}

Directly using the result \eqref{main_result}, it is not hard to calculate the four-point function of the scalar primary operator $\sigma$ in the critical fermionic theory. 
The diagrams that contribute to the four-point function of $\sigma$ are  shown in Figure \ref{fig:critical_fermions}.
\begin{align}
&&\langle \hat{\sigma}(-q_1)\hat{\sigma}(-q_2)\hat{\sigma}(-q_3)\hat{\sigma}(-q_4)\rangle=N^4 \tilde{\lambda}_{q.b}^4G(-q_1)G(-q_2)G(-q_3)G(-q_4)\nonumber \\ && (\langle  J^{(0)}(-q_1)J^{(0)}(-q_2)J^{(0)}(-q_3)J^{(0)}(-q_4) \rangle + \nonumber \\ && G(q_1+q_2)(\langle  J^{(0)}(-q_1)J^{(0)}(-q_2)J^{(0)}(q_1+q_2) \rangle \langle  J^{(0)}(-q_3)J^{(0)}(-q_4)J^{(0)}(q_3+q_4) \rangle + \nonumber \\ &&\lambda_6(\langle  J^{(0)}(-q_1)J^{(0)}(-q_2)J^{(0)}(q_1+q_2) \rangle +\langle  J^{(0)}(-q_3)J^{(0)}(-q_4)J^{(0)}(q_3+q_4) \rangle)+\lambda_{6}^2)) + \text{ permutations}
\label{eq:42}
\end{align}

The critical fermionic theory is dual to the non-critical bosonic theory in the limit $\lambda_f \rightarrow 1$ and $\lambda_6=0$. In this limit, there are two surviving terms in \eqref{eq:42}, the first two diagrams of Figure \ref{fig:critical_fermions}. Using \eqref{main_result} and the fermionic three point function \cite{GurAri:2012is} we find
\begin{align}
&\langle \hat{\sigma}(-q_1)\hat{\sigma}(-q_2)\hat{\sigma}(-q_3)\hat{\sigma}(-q_4)\rangle=-\frac{256 \pi^4 \lambda_{f}^4\cot^8 \left(\frac{\pi \lambda_f}{2} \right)}{|q_1||q_2||q_3||q_4|} \times \nonumber \\ &  \Big(\frac{2\sec^2\left(\frac{\pi \lambda_f}{2}\right) \tan \left(\frac{\pi \lambda_f}{2}\right) }{\pi \lambda}F(q_1,q_2,q_3,q_4) +H(q_1,q_2,q_3,q_4) \nonumber \\ & +\frac{N_f\tan^3 \left(\frac{\pi \lambda_f}{2}\right)}{\pi \lambda_{f}} \left(\frac{1}{|q_1+q_2|}+\frac{1}{|q_1+q_3|}+\frac{1}{|q_3+q_3|} \right) \Big)
\\ &
=-\frac{256 \pi^4 \lambda_{f}^4}{|q_1||q_2||q_3||q_4|} \cot^4 \left(\frac{\pi \lambda_{f}}{2} \right)\frac{|\lambda_b|}{|\lambda_f|}\left(F(q_1,q_2,q_3,q_4)+N_f\left(\frac{1}{2|q_1+q_2|}+\frac{1}{2|q_1+q_3|}+\frac{1}{2|q_3+q_3|}\right)\right)
\label{eq:43}
\end{align}
%FIX THIS EQUATION
In the last step we have applied the limit $\lambda_f \rightarrow 1$ and used \eqref{duality} to express the answer in terms of $|\lambda_b|$, anticipating the comparison in the next section. We have also used the fact that $H(q_1,q_2,q_3,q_4) \sim \tan^2\left(\frac{\pi \lambda_f}{2}\right)$ to eliminate it from \eqref{eq:43}. Using equation \eqref{norm}, this expression can be compared to the four-point correlator of $\hat{J}^{(0)}_b$ in the bosonic theory which we calculate in the next section.

\section{Comparison to the Bosonic Theory}
\label{bosonic_theory_section}
\subsection{Non-Critical Bosonic Theory}
In the free theory, we have
\begin{align}
\langle J^{(0)}(-q_1)J^{(0)}(-q_2)J^{(0)}(-q_3)J^{(0)}(-q_4) \rangle && 
 \\&& \hspace{-30mm}= N\int \frac{d^3k}{(2\pi )^3}\frac{1}{k^2(k+q_1)^2(k+q_1+q_2)^2(k+q_1+q_2+q_3)^2}
 \label{eq:32}
 \end{align}
 We solve \eqref{eq:32} in the limit $q_\pm=0$. The integral evaluates to 
 \begin{align}
&&\frac{N}{2}\Big(\frac{(q_2^2+q_3^2+q_1q_2+q_1q_3+q_2q_3)}{|q_1|q_2q_3(q_1+q_2+q_3)(q_1+q_2)(q_1+q_3)(q_2+q_3)}+  (q_1\rightarrow q_2,q_3) \nonumber \\ &&+ \frac{q_1q_2+q_1q_3+q_2q_3}{q_1q_2q_3|q_1+q_2+q_3|(q_1+q_2)(q_1+q_3)(q_2+q_3)}\nonumber \\ &&- \frac{(q_1+q_3)(q_2+q_3)}{(q_1q_2q_3(q_1+q_2+q_3)|(q_1+q_2)|(q_1+q_3)(q_2+q_3))}+(q_1 \leftrightarrow q_2 \leftrightarrow q_3))\Big)
\label{eq:33}
\end{align}
where the arrows in the numerator imply symmetric terms on replacing $q_1$ with $q_2$ and $q_3$.
\begin{figure}
\centering % \begin{center}/\end{center} takes some additional vertical space
\includegraphics[scale=0.7]{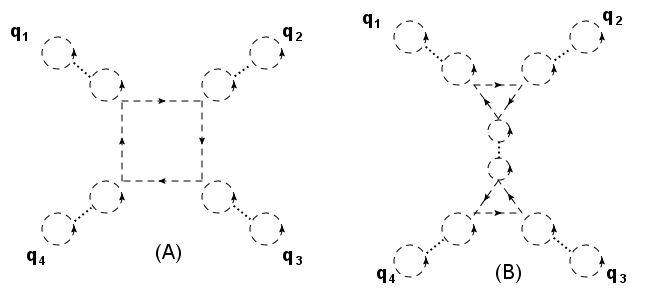}
\caption{\label{fig:i:7} Correction to the critical scalar four correlator.}
\end{figure}

Using the normalisation relation 
\begin{equation}
\hat{\sigma}=-4 \pi \lambda_f \cot \left(\frac{\pi \lambda_f}{2} \right) J^{(0)}_b \label{norm}
\end{equation}
which follows from comparing the two point functions of the scalar primaries in both theories, it is easy to see that this result matches the R.H.S of \eqref{eq:43} -- i.e., that $$ \langle \sigma(-q_1) \sigma(-q_2) \sigma(-q_3) \sigma (-q_4)\rangle = \left(-4\pi \lambda_f \cot \left(\pi \lambda_f /2 \right) \right)^4 \langle J^0_b(-q_1) J^0_b(-q_2)  J^0_b(-q_3)  J^0_b(-q_4)  \rangle,$$ thereby verifying the duality between the critical fermionic theory and the non-critical bosonic theory (for our restricted choice of external momenta).

\subsection{Critical Bosonic Theory}
%CRITICAL BOSONIC THEORY
Next we turn to the four point correlator at the critical fixed point of the theory.  This is accomplished by adding a double trace term to the scalar action. The vertex is given by $-\frac{\lambda_4}{N}(\phi^\dagger \phi)^2$. We next flow to the IR limit with the IR scalar mass zero by tuning $\lambda_4 $ to infinity. The scalar propagator does not get a finite correction from this deformation. The divergent terms can be subtracted by a mass counterterm. Two and three point correlators in the critical theory were discussed in \cite{Aharony:2012nh, GurAri:2012is}.

The four point correlator receives a correction from the $\lambda_4$ deformation in two diagrams, one with a four point diagram Figure~\ref{fig:i:7}(A) and one with a double three-point diagram Figure ~\ref{fig:i:7}(B).
 Diagram ~\ref{fig:i:7}(B) can be evaluated as
 \begin{eqnarray}
&&
\langle J^{(0)}(-q_1)J^{(0)}(-q_2)J^{(0)}(-q_3)J^{(0)}(-q_4) \rangle_{\lambda_4}^B= \nonumber 
\\ && 
\langle J^{(0)}(-q_1)J^{(0)}(-q_2)J^{(0)}\rangle_{\lambda_4=0}\langle J^{(0)}(-q_3)J^{(0)}(-q_4)J^{(0)}\rangle_{\lambda_4=0} \nonumber 
\\ &\times & 
  \prod \limits_{q=q_1,q_2,q_3,q_4,q_1+q_2} \sum \limits_{n=0}^{\infty} \left(-\frac{\lambda_4}{N_b}\langle J^{(0)}(-q)J^{(0)}\rangle_{\lambda_4=0}\right)^n \nonumber \\
  && + \text{ 3 permutations: } (q_1+q_2 )\rightarrow (q_1+q_3),(q_1+q_4)  
\label{eq:34}
 \end{eqnarray}
 The values of $\langle J^{(0)}(-q_1)J^{(0)}\rangle_{\lambda_4=0}$ and $\langle J^{(0)}(-q_1)J^{(0)}(-q_2)J^{(0)}\rangle_{\lambda_4=0}$ are given in  \cite{Aharony:2012nh, GurAri:2012is}. We are interested in the free scalar theory hence we take the momentum dependence as 

 \begin{gather}
 \langle J^{(0)}(-q_1)J^{(0)}\rangle_{\lambda_4=0}=N_b\frac{1}{8|q|} 
 \label{eq:35}
 \\\langle J^{(0)}(-q_1)J^{(0)}(-q_2)J^{(0)}(q_1+q_2)\rangle_{\lambda_4=0}=N_b\frac{1}{4|q_1||q_2||q_1+q_2|}
 \label{eq:36}
 \end{gather}
 
 \eqref{eq:35} and \eqref{eq:36}can be substituted in \eqref{eq:34} to give 
 \begin{align}
 \hspace{-5mm}\frac{-N_b\lambda_4}{8|q_1||q_2||q_3|q_4||q_1+q_2|^2} \prod \limits_{q_1,q_2,q_3,q_4,q_1+q_2}\frac{8|q_1|}{\lambda_4 \left(1+\frac{16|q_1|}{\lambda_4}\right)}+\left((q_1+q_2 )\rightarrow (q_1+q_3),(q_1+q_4)\right)
 \label{eq:37}
 \end{align}
 We define $\tilde{J}^{(0)}=\lambda_4J^{(0)}$ as the scalar operator at the critical fixed point. In the IR limit, taking $ \lambda_4\rightarrow \infty$ expanding the denominator and keeping the leading term we get 
 \begin{equation}
 \langle\tilde{J}^{(0)}(-q_1)\tilde{J}^{(0)}(-q_2)\tilde{J}^{(0)}(-q_3)\tilde{J}^{(0)}(q_1+q_2+q_3)\rangle_{\lambda_4}^B=-\frac{8^4N_b}{2}\left(\frac{1}{|q_1+q_2|}+\frac{1}{|q_1+q_3|}+\frac{1}{|q_2+q_3|}\right)
 \label{eq:38}
 \end{equation}
Figure  ~\ref{fig:i:7}(A) turns out to be
  \begin{eqnarray}
&& \hspace{-5mm}\langle J^{(0)}(-q_1)J^{(0)}(-q_2)J^{(0)}(-q_3)J^{(0)}(-q_4)\rangle_{\lambda_4}^A  \nonumber \\ && \hspace{-5mm}
= \langle J^{(0)}(-q_1)J^{(0)}(-q_2)J^{(0)}(-q_3)J^{(0)}(-q_4)\rangle_{\lambda_{4=0}} \hspace{-.5cm} \prod \limits_{q=q_1,q_2,q_3,q_4}\sum \limits_{n=0}^{\infty} (-\frac{\lambda_4}{N_b}\langle J^{(0)}(-q)J^{(0)}\rangle)^n
 \label{eq:39}
  \end{eqnarray}
 We can use \eqref{eq:33} and use the same methods to get 
 \begin{eqnarray}
 &&\langle\tilde{J}^{(0)}(-q_1)\tilde{J}^{(0)}(-q_2)\tilde{J}^{(0)}(-q_3)\tilde{J}^{(0)}(q_1+q_2+q_3)\rangle_{\lambda_4}^A=8^4N_b \nonumber \\ &&
 (\text{sign}[q_1](q_2^2+q_3^2+q_1q_2+q_1q_3+q_2q_3)+  (q_1\rightarrow q_2,q_3))+(\text{sign}[q_1+q_2+q_3](q_1q_2+q_1q_3+q_2q_3))\nonumber \\ && \frac{-(\text{sign}[q_1+q_2](q_1+q_3)(q_2+q_3)+(q_1 \leftrightarrow q_2 \leftrightarrow q_3))}{(q_1+q_2)(q_1+q_3)(q_2+q_3)}\text{sign}[(q_1q_2q_3(q_1+q_2)]
 \label{eq:40}
 \end{eqnarray}
  
 Adding \eqref{eq:38} and \eqref{eq:40} to obtain $\langle \tilde J^{(0)}(-q_1) J^{(0)}(-q_2)J^{(0)}(-q_3)J^{(0)}(-q_4) \rangle $, it is easy to see that the momentum dependence equates to the non-critical free fermion correlator \eqref{eq:41}.
Employing normalisation $\tilde{J}_b^{(0)} = 4\pi \lambda_b J^{(0)}_f$ obtained in Section \ref{sec:normalisation} and applying the limit $\lambda_f \rightarrow 1$ to \eqref{main_result} we find 
 \begin{equation}
\langle \tilde{J}_b^{(0)}(-q_1) \tilde{J}_b^{(0)}(-q_2)\tilde{J}_b^{(0)}(-q_3)\tilde{J}_b^{(0)}(-q_4) \rangle = (4\pi\lambda_b)^4 \langle J_f^{(0)}(-q_1) J_f^{(0)}(-q_2)J_f^{(0)}(-q_3)J_f^{(0)}(-q_4) \rangle  \end{equation} 
thereby verifying the duality between the critical bosonic theory and the non-critical fermionic theory for our restricted choice of external momenta.
  
\section{Discussion}

The main result of the paper is \eqref{main_result}, an explicit expression for the four-point function of the scalar primary in a particular limit of external momenta for both the non-critical fermionic theory. We also calculated the four-point function in the critical fermionic theory, and compared to critical and non-critical free bosons, providing an independent confirmation of the bosonization duality introduced in section \ref{sec:intro} at the level of four-point functions.

Our calculations crucially relied on the off-shell exact ladder diagram \eqref{ladder1}-\eqref{ladder4} together with \eqref{eq:19}. It is relatively straightforward to solve the resulting integral equations for the case when $q_\pm=0$. However, if we could generalise the calculation above to the case $q_\pm \neq 0$, we would be able to calculate four-point functions with arbitrary momenta. More importantly, the off-shell ladder diagram is also required for calculating $1/N$ corrections (and $M/N$  corrections in a bifundamental theory, see \cite{Guru:2014ad}) to all orders in $\lambda$. We hope to return to this off-shell ladder diagram in the future. We note that the \textit{on-shell} four point function is calculated to all orders in \cite{Jain:2014nza}, and its supersymmetric generalization \cite{Inbasekar:2015tsa}.

The theories we study possess a slightly-broken higher spin symmetry. In \cite{Maldacena:2011jn, Maldacena:2012sf}, two different classes of large $N$ field theories with a slightly broken higher spin symmetry were found to exist -- "quasi-boson" and "quasi-fermion" theories. The quasi-fermion theory depends on two parameters, $\tilde{N}$ and 
$\tilde{\lambda}$.  The quasi-boson theories depend on three parameters, $\tilde{N}$ and 
$\tilde{\lambda}$ and $a_3$. The first two parameters essentially correspond to the rank of the gauge group and the 't Hooft coupling $\lambda=\frac{N}{k}$ (in a microscopic description) and the third parameter corresponds to the $\phi^6$ triple-trace coupling which is exactly marginal in the large $N$ limit of the bosonic theory. 

While three point functions in conformal field theories are severely constrained by purely kinematic considerations, four-point functions are determined only up-to an undetermined function of two conformal cross-ratios. In particular, conformal invariance restricts the four point function of scalars $J^0$, with scaling dimension $\Delta$ to the   form:
\begin{equation}
\langle J^0(x_1) J^0(x_2) J^0(x_3) J^0(x_4) \rangle = \frac{1}{x_{12}^{2\Delta} x_{23}^{2\Delta}}f(u,v)
\end{equation}
where $f$ is any function of the conformally invariant cross-ratios $u= \frac{x_{12}^2 x_{34}^2}{x_{13}^2 x_{24}^2}$ and $v = \frac{x_{14}^2 x_{32}^2}{x_{13}^2 x_{42}^2}$, satisfying $f(u,v)=\left(u/v\right)^\Delta f(v,u)$. However, dynamically, all correlation functions in a conformal field theory are (in principle) uniquely determined by the three-point functions (i.e., the operator-algebra) and scaling dimensions of the primary operators in the theory via bootstrap arguments.\footnote{We note that the planar four-point function calculated here is expected to receive contributions from the three-point functions of multi-trace as well as single-trace operators. Previous works have checked the duality at the level of planar three-point functions of single-trace operators but not multi-trace operators, so the calculation of this paper provides an independent check of the duality. We thank an anonymous referee for clarifying this point.} It would be interesting to study the four-point functions in theories with a slightly broken higher spin symmetry via bootstrap arguments. We hope the explicit results for Chern-Simons vector models here derived here would be useful in such a program.

\textbf{Acknowledgements:}
The authors thank Shiraz Minwalla and V. Umesh for reading a draft for the paper. SP acknowledges support of a DST Inspire Faculty Award.
SP would like to thank the Centre for High Energy Physics, IISc Bangalore and IISER Pune for hospitality when part of this work was completed and CMS, Durham University and the Rudolf Peierls Centre for Theoretical Physics, Oxford University for hospitality during the final stages of this work.

\appendix

\section*{Appendix A: Exact vertices from the ladder diagram}
As a check on our calculation, the ladder diagram can be utilised to evaluate the exact $J^{(0)}$  vertex derived earlier in \cite{ GurAri:2012is}.

For $J^{(0)}$, $f_I^0$ and $f_-^0$ can be seen from Fig.~\ref{fig:i:2} with the contracted vertex on the right as the free scalar vertex. Subsequently, it can be written as
\begin{equation}
N\frac{f_-^0}{(p-r)^+} =-i \lambda \int \frac{d^3k}{(2\pi)^3}\frac{ k^+(2i\lambda k_s-q_3)}{k^2(k+q_3)^2}\frac{1}{(p-k)^+}  
\end{equation}
\begin{equation}
N \frac{f_I^0}{(p-r)^+} =-i \lambda \int \frac{d^3k}{(2\pi)^3}\frac{2k_s^2(1-\lambda^2)}{k^2(k+q_3)^2}\frac{1}{(p-k)^+}  
\end{equation}
Using eq.\eqref{ladder1}-\eqref{ladder4}, this immediately yields
\begin{equation}
J^{(0)}(q_3)-I=\frac{e^{-2i\lambda \arctan[\frac{2p}{q_3}]}-e^{-2i\lambda \arctan[\frac{2\Lambda}{q_3}]}}{1+e^{-2i\lambda \arctan[\frac{2\Lambda}{q_3}]}} I + \frac{q_3}{2 p^+}\frac{(q_3-2i\lambda p-(q_3+2i\lambda p)e^{-2i\lambda \arctan[\frac{2p_s}{q_3}]})}{1+e^{-2i\lambda \arctan[\frac{2\Lambda}{q_3}]}}\gamma^+
\end{equation}
which on the substitution \eqref{eq:27} gives us the exact $J^{(0)}$ vertex.

\bibliography{CSBib}
\bibliographystyle{JHEP}

\end{document}